\documentclass[10pt,a4paper,]{article}
\usepackage{amsmath,amssymb,amsfonts,amsthm}
\usepackage[latin1]{inputenc}
\usepackage[english]{babel}
\usepackage{graphics}
\usepackage{amsfonts}

\usepackage[cmtip,all]{xy}
\usepackage[all]{xy}
\xyoption{all}
\xyoption{2cell}

\mathcode`\<="4268 
\mathcode`\>="5269 
\mathchardef\gt="313E 
\mathchardef\lt="313C 

\theoremstyle{definition} \newtheorem{Fell bundle geometry}[subsection]{Definition}

\theoremstyle{definition} \newtheorem{tangent field}[subsubsection]{Definition}
\theoremstyle{definition} \newtheorem{tangent space}[subsubsection]{Definition}
\theoremstyle{definition} \newtheorem{tangent sheaf}[subsubsection]{Definition}
\theoremstyle{definition} \newtheorem{cotangent sheaf}[subsubsection]{Definition}
\theoremstyle{definition} \newtheorem{sheaf opp}[subsubsection]{Definition}
                          \newtheorem{spaceoid}[subsection]{Definition}

                          \newtheorem{sheaf Cl}[subsection]{Definition}

                          \newtheorem{Constraint corollary}[subsection]{Corollary}
                          \newtheorem{General D}[subsubsection]{Definition}

\theoremstyle{definition}     \newtheorem{comments on sheaf}[subsubsection]{Comment}
                                \newtheorem{example FB}[subsection]{Example}
                          \newtheorem{example 2 FB}[subsection]{Example}
                          \newtheorem{example 3 FB}[subsection]{Example}
                          \newtheorem{example FBG}[subsection]{Example}
                          \newtheorem{example FBG nc}[subsection]{Example}
                          \newtheorem{Kasparov comment}[subsection]{Comment}
                          \newtheorem{compact comment}[subsection]{Comment}

                          \newtheorem{comment almost com}[subsection]{Comment}
                          \newtheorem{explanatory remark}[subsection]{Remark}
                          \newtheorem{comment flat}[subsection]{Comment}
                          \newtheorem{bounded remark}[subsection]{Remark}
                          \newtheorem{Hilbert space comment}[subsection]{Comment}

\theoremstyle{definition} 
\theoremstyle{remark}     \newtheorem{comment on metric}[subsubsection]{Comment}
\theoremstyle{definition} \newtheorem{Fell bundle D}[subsubsection]{Definition}
\theoremstyle{definition}     \newtheorem{not inner}[subsubsection]{Remark}
\theoremstyle{definition}     \newtheorem{junk}[subsubsection]{Comment}
\theoremstyle{definition}     \newtheorem{comment line}[subsubsection]{Comment}
\theoremstyle{definition}     
\theoremstyle{definition}     
\theoremstyle{definition}     \newtheorem{Comment:}[subsection]{Comment}
\theoremstyle{definition} \newtheorem{two point space}[subsection]{Example}
\theoremstyle{definition} 
\theoremstyle{definition} 
\theoremstyle{definition} \newtheorem{Dirac operator}[subsubsection]{Definition}

\theoremstyle{definition}     \newtheorem{sections}[subsubsection]{Remark}
\theoremstyle{remark}     \newtheorem{metric remark}[subsubsection]{Remark}

\theoremstyle{plain}  \newtheorem{main theorem}[subsection]{Theorem}

\theoremstyle{definition} \newtheorem{Corollary of main theorem}[subsection]{Corollary of (a)}

\begin{document}

\title{Categorified Noncommutative manifolds}

\author{R.A. Dawe Martins \\\\ Centro de An\'alise Matem\'atica,\\ Geometria e Sistemas
Din\^amicas \\ Departamento de Matem\'atica \\ Instituto Superior T\'ecnico \\ Av. Rovisco Pais \\
1049-001 Lisboa \thanks{Email: rmartins@math.ist.utl.pt. Research supported by Fundaç\~ao para a Ci\^encias e a Tecnologia (FCT)
including program POCI 2010/FEDER.}}

\date{\today}

\maketitle

\begin{abstract}
We construct a noncommutative geometry with generalised `tangent bundle'
from Fell bundle $C^*$-categories ($E$) beginning by replacing pair groupoid objects (points) with
objects in $E$. This provides a categorification of a certain class of real spectral triples where
the Dirac operator $D$ is constructed from morphisms in a category. Applications for physics
include quantization via the tangent groupoid and new constraints on $D_{\mathrm{finite}}$ (the
fermion mass matrix). \end{abstract}

\section{Introduction}

This contribution is an excerpt from the paper ``Some constructions in Category
theory and Noncommutative geometry'' and is based on a short talk. For this reason many of the
preliminary details, examples and the proofs have had to be omitted from the current paper.

According to  Connes and Chamseddine \cite{sap} the world is a product of four dimensional spacetime
and a noncommutative manifold capturing the charges and chiralities of the particles of the standard
model. The algebra of the total space is a tensor product $C^{\infty}(M_4) \otimes A_F$ while the
multiplication of the two Riemannian spin geometries, formulated as real spectral triples, is
product over $K$-cycles. We suggest that if this ``almost commutative'' spectral triple be the
correct point of view, then quantum gravity tools and ideas on spacetime might be first generalised
to the noncommutative factor and then extended to the total space.

In some ways  it may be surprising that the noncommutative standard model is not already quantum
from first principles. It is afterall based on a noncommutative algebra. As a fully geometrical
theory (the action depends only on the eigenvalues of the Dirac operator and so it is pure geometry
and diffeomorphism invariant) its quantisation would involve quantum gravity in some sense
\cite{GAN}. Moreover, Connes has constructed an analogue of general relativity for the discrete
space where gravity is the ``Higgs pseudoforce'' with equivalence principle, Einstein's equations
and the Higgs field as a connection (for details see for example \cite{forces}). The fermion masses
can be viewed as coming from the work done against the Higgs force as a particle is parallel
transported between chiralities. Even though the model does require tuning and data input, it is a
predictive theory and as described above, is much more than a repackaging of the standard model.
Aastrup, Grimstrup and Nest \cite{GAN},\cite{GAN2} already established a link between
Connes's noncommutative geometry and loop quantum gravity. This
paragraph is an attempt to describe the intersection of their motivation with that for this paper.

Where relativity  is involved, the formula ``classical mechanics + noncommutativity = quantum
mechanics'' is no longer useful and in the case of gravity in particular, Crane has argued that
``general relativity + category theory = quantum gravity'' (\cite{qg}, \cite{category qg}). We
suggest therefore that since the noncommutative standard model involves general relativity on the
total space perhaps it is category theory that is missing, that is, one of the reasons why it is a
not a quantum theory. The main point of this paper is to begin to develop a categorification of the
notion of real spectral triple and to describe some mathematical and physical consequences of doing
that. For this we use Fell bundle $C^*$-categories. Bertozzini, Conti and Lewkeeratiyutkul
\cite{Paolo2} \cite{Paolo} have already brought spectral triples into the subject of category
theory.

A second branch of this paper on a generalisation of the tangent
groupoid to noncommutative manifolds, that is, real spectral triples. Again we will again need the
concept of a Fell bundle and from a geometrical point of view as well as from a categorical point
of view. Our discussion is different in nature from the usual generalisations of the tangent
groupoid because instead of choosing a favourite space to replace $\mathbb{R}^n$ and then working
out the Moyal quantization to attain a noncommutative algebra, we already have noncommutativity of
functions and therefore the algebra of observables will automatically be noncommutative as well.
Instead, the nature of this generalisation is to formulate an algebraic ``tangent bundle'' for real
spectral triples, which we will call ``sheaf of tangent fields''.

Note that the categorification process presented here necessarily leads \emph{directly} to
quantization of particle mechanics only in the case of a commutative manifold.


The Dirac operator  for a spectral triple is defined without reference to a tangent bundle. This is
necessary because real spectral triples are supposed to give a solely algebraic characterisation of
Riemannian manifolds whereas tangent vectors involve the notions of points and angles. However, in
order to apply the tangent groupoid to attain an algebra of observables for a particle system on a
space, that space needs a (co)tangent bundle - a phase space. To overcome this we develop an
algebraic notion to replace that of tangent bundle, one of ``tangent sheaf''. The mathematical
source being a Fell bundle $C^*$ category together with ideas from the tangent groupoid.

Another motivation  for an explicit construction of $D$ from a cotangent sheaf is that
$D_{\mathrm{finite}}$ for the standard model is an ad hoc insertion to the theory because it
involves the fermion mass matrix. We show by example that the current application of category
theory can constrain $D_{\textrm{finite}}$ for example ensuring the photon and the gluons remain
massless. Therefore, one more aim of this paper is to evidence that the mathematics of Fell bundle
$C^*$-categories can provide the necessary enrichment of real spectral triples to constrain
$D_{\mathrm{finite}}$ beyond the noncommutative geometry axioms due to their algebraic notions such
as groupoids, Morita equivalence bimodules and category theory. In addition, they are orientable by
definition and their fibres are not necessarily isomorphic. This leads to a new description of a
geometry on a noncommutative manifold, which is more explicit but still point-free.

We begin by putting  a geometry on a Fell bundle $E$ over a principal groupoid. That means, we
define for $E$ a counterpart of Riemannian metric and distance function, sheaf of `tangent fields'
and a Dirac operator. Since there are two noncommutative algebras we have to check 2
correspondences to classical limits. These spaces are supposed to be noncommutative generalisations
of manifolds and therefore they should be real spectral triples. We check that Fell bundle
geometries as we define them are examples of real spectral triples. Since they are
$C^*$-categories, they can  be considered a categorification of real spectral triples up to the
breadth of the class of real spectral triples that can be constructed as Fell bundle geometries. As
a result, the Dirac operator attains the status of a field of morphisms in a category and Fell
bundle geometries can be described as deformed not necessarily commutative geometries with explicit
construction of $D$ and involving a groupoid notion of tangent sheaf.

The idea that a good  notion of generalised tangent bundle can be formalised with groupoids is not
new. There is of course the tangent groupoid itself (Connes and Landsman, see for example
\cite{Connes's book}) and Debord and Lescure \cite{dl} have also worked on this.

\section{Preliminaries}

\subsection{\textit{Fell bundle $C^*$-categories}}

\begin{spaceoid}[Fell bundle $C^*$-category]  A Fell bundle $C^*$-category $E$ is a saturated
unital Fell bundle over a principal groupoid (or equivalence relation). A one-dimensional Fell
bundle $C^*$-category is called a `topological spaceoid' \cite{Paolo}.
\end{spaceoid}

To see that the above is a $C^*$-category, one views the elements of the Morita
equivalence bimodules as morphisms and the $C^*$- algebras over the groupoid
units as objects. These have to be unital algebras to satisfy the category theoretic unit law and
finally the multiplication in the Fell bundle is associative by definition. This is a
small subcategory of the category of Banach spaces together with a *-functor $\pi : E \rightarrow
G$. The practical implication of $E$ being \textit{saturated} ($E_{\gamma_1}.E_{\gamma_2}$ is total
in $E_{\gamma_1 \gamma_2}$ for all $(\gamma_1, \gamma_2)~\in~\Gamma^2$) is that the latter are
Morita equivalence bimodules. Note that finite dimensional Fell bundles are saturated. As is true
for any $C^*$-category, it can be represented on a concrete $C^*$-category, (that is, a small
subcategory of Hilb\footnote{objects are Hilbert spaces and morphisms are the bounded linear maps
between them.}). See \cite{LRG}, \cite{Paolo}. For the definition of Fell bundle see \cite{fbg}.

\subsection{\textit{Fluctuations of the metric}}

In \cite{forces} Sch\"ucker explains that while Einstein derived general relativity from Riemannian
geometry, Connes extended this to noncommutative geometry and as a result the other three
fundamental forces emerged with the gauge and Higgs fields as fluctuations of the metric. The
`almost commutative' spectral triple of the noncommutative standard model includes a commutative
space-time factor and a finite noncommutative factor. The latter is reminiscent of Kaluza-Klein
internal space, but in this case it is 0-dimensional. Connes' encodes the metric data in the Dirac
operator \cite{gravity} and his procedure for unification starts by describing the diffeomorphisms
giving rise to the equivalence principle as the spinor lift of the automorphisms of the algebra
transforming the Dirac operator.  The arising space of fluctuated Dirac operators $D^f$  (\cite{ncg
and sm}) defines the configuration space of the spectral action:-

\begin{equation} \label{fluctuations}
  D^f = \sum_{\textrm{finite}} r_j L (\sigma_j) D L(\sigma_j)^{-1}, ~~r_j \in \mathbb{R}, ~~ \sigma_j \in \textrm{Aut}(A)
\end{equation}

where $L$ is the double valued lift of the automorphism group to the spinors. For the calculation of
the spectral standard model action see \cite{sap}. The result is the general form of the Dirac
operator with arbitrary curvature and torsion:

\begin{equation} \label{general D}
 D =  \sum_i c_i \Big(\frac{d}{dx}_i + \omega_i \Big)
\end{equation}

The  `almost commutative' algebra of the noncommutative standard model comprises two factors,
 $A = C^{\infty}(M) \otimes A_{\mathrm{finite}}$ and `fluctuating' the Dirac operator that probes
 $M$ in the finite  space algebra $A_F$, Connes obtains the standard model gauge
fields. In this noncommutative case, this means replacing the spinor lift in the above formula with
the unitaries of $A_{\mathrm{finite}}$. Finally, fluctuating the Dirac operator
$D_{\mathrm{finite}}$ that probes the finite space in $A_{\mathrm{finite}}$ gives rise to the Higgs
field. This gives the Higgs field an interpretation as a gravitational connection on an additional
`dimension'.

\section{Fell bundle geometries}

A Fell  bundle geometry is intended to provide a category theoretic `switch-of-focus' from the
deformed Riemannian geometry defined by the generators of the noncommutative algebra $C^*(M \times
M)$, directly generalisable to the case of a noncommutative manifold. Fell bundle geometries can be
described as deformed not necessarily commutative geometries with an explicit construction of $D$
and possessing a generalised `tangent bundle' or tangent sheaf. Having such a notion is supposed to
make them a receptacle for the application of tangent groupoid quantisation to noncommutative
spaces. Since the algebra is noncommutative from the starting point, the work to be done is not in
deforming a commutative algebra into a noncommutative one, but in developing an algebraic
description of the deformed geometry. What we mean by \textit{deformed geometry} is one defined by
the generators of the noncommutative algebra of observables.  In this way, the groupoid
$G=M \times M$ is the deformed tangent bundle (\cite{dl}). The topology of the manifold $M$ is
unchanged because it is identified with $G_0$, which is the same for all $\hbar$. It is only the
tangent bundle and hence the geometry that changes.

\begin{Fell bundle geometry} A \textit{Fell bundle geometry}   $((E \otimes E^{opp}, \pi, G), D, H)$
or $(E,D)$ is a tensor product Fell bundle $C^*$-category $(E \otimes E^{opp}, \pi, G)$, where $G_0$
is a compact Riemannian manifold $M$, together with the following structures to provide a ``geometry'':

(i) a `tangent sheaf':- an algebraic formulation of generalised tangent bundle, (see
definitions below);

(ii) analogues of Riemannian metric and Riemannian distance, (see definition 3.11.1 in the main
paper: "Some constructions in Category theory and Noncommutative geometry");

(iii) a Dirac operator $D$, (see definition below)

and where the algebra $C^*(E^0) \otimes C^*(E^{opp,~0})$ or $A \otimes A^{opp}$ is faithfully
represented on a Hilbert space $H$, which is a finite projective module with
$\mathbb{Z}_2$-grading $\chi$ and with reality structure $J$. The data $(A^{opp}, A,
F=\mathrm{sign} D, \chi, H)$ is a Kasparov module.

Unless otherwise stated, $M$ will be spin in which case $DJ=JD$ and there is a Fell bundle
$C^*$-category $Cl$ over $G$ whose restriction to $G_0$ is the complexified Clifford bundle over
$M$, which we denote $Cl^0$. Otherwise, $DJ=-JD$. If $A$ is
noncommutative then $E$ and $Cl$ will denote the same Fell bundle.

\end{Fell bundle geometry}

\begin{example FBG}[Commutative Fell bundle geometries] Consider the special case when $A$ is
commutative. Let $M$ be a Riemannian spin manifold. We associate to $M$ a Fell bundle geometry
$(E,D)$ as follows. $E$ and $E \otimes E^{opp}$ are complex line bundles over the Lie groupoid $G =
M \times M$, that is to say, a topological spaceoid \cite{Paolo}. The algebra of sections $A$ of the
$C^*$-algebra bundle $E^0$ is identified with the $C^*$ completion of the complex valued algebra of
infinitely differentiable functions on $M$. $H$ is the Hilbert space of square integrable sections
of the spinor bundle associated with the Clifford bundle $Cl^0$.  The geometry of $(E,D)$ is given
by (i) to (iii).                     \end{example FBG}

\begin{example FBG nc}[Noncommutative Fell bundle geometries]  A noncommutative Fell bundle geometry
is  a finite dimensional Fell bundle $E$ over a
discrete pair groupoid $G$ together with (i) to (iii) above. Note that $M$ is a
0-dimensional topological space. Note also that the fibres of $E$ are not necessarily isomorphic
unlike in the case of commutative $(E,D)$. The objects are finite dimensional simple algebras:-
$M_{n_i}(\mathbb{C})$ where $i$ is an index set on the groupoid unit space. Since
$G$ is discrete and $M$ is compact, $H$ must be finite dimensional, $H = \mathbb{C}^m$ where $m =
\sum_i n_i$.

Note that for noncommutative $(E,D)$ the manifold that we associate
the Fell bundle to is not $M$ but  the intuitively larger virtual space encoded by $A$; since the
fibres are not necessarily copies of $\mathbb{C}$, but in general are the larger algebras
$M_n(\mathbb{C})$, the map $\pi : E_0 \rightarrow G_0$ is not necessarily defined by the Gelfand
functor.
      \end{example FBG nc}



\begin{Kasparov comment} In  order to make sure that the spectral triple axiom of Poincar\'e duality
is satisfied (\cite{gravity}) we impose that $(A, ~ A^{opp}, ~ \mathrm{sign}D, ~ H, ~ \chi )$ be a
Kasparov module, so that we have the isomorphisms $K_i(A) \cong K^i(A^{opp})$ where $i=0$ or 1. In
other words, we set $(E,D)$ to have Poincar\'e duality by definition. In doing this, we can say that
to some extent we are copying in the topological and analytical properties from a spectral triple
into this definition, while constructing new geometrical properties for $E$. Not relevant in the
case of finite dimensional $H$, the other conditions on $A$ and $H$ from noncommutative geometry
without explicit involvement of the spectrum of $D$ are smoothness of coordinates and absolute
continuity (\cite{gravity}).
                         \end{Kasparov comment}

\begin{compact comment} Note that the reason for restricting the scope to compact $M$ or $G_0$ was
in order to accommodate the fact that we are working with $C^*$-categories and therefore the object
$C^*$-algebras must be unital.
                       \end{compact comment}

\begin{comment almost com} One could also construct an `almost commutative' Fell bundle geometry as
a $K$-cycle product of a commutative with a noncommutative Fell bundle geometry.
                           \end{comment almost com}

\begin{Hilbert space comment} The structure group of the Clifford bundle is $\mathrm{End}(Cl^0)$ and
the Hilbert space from the Gelfand-Naimark-Segal construction comes from left multiplication of
elements of $Cl^0$, which give rise to operators in    $\mathrm{End}(Cl^0)$. The Hilbert
space that a Fell bundle $C^*$-algebra acts on can be constructed from the GNS construction
and therefore the Hilbert space $L^2(M;S)$ should be closely related to the Hilbert space one would
normally associate to a Fell bundle.
                                    \end{Hilbert space comment}

\subsection{\textit{Tangent sheaf}}

It is well known that point-set  topology can be replaced by analysis with no loss of information,
and spectral triples are based on the paradigm that geometry can have an equivalent algebraic
description with commutative geometry as a special case.  With this section we intend to formalise
our view that a Fell bundle geometry as a noncommutative Riemannian manifold (spectral triple)
suggests a `space' comprising objects in a category and a `tangent bundle' or rather a tangent
sheaf of continuous fields of morphisms in the category. This is similar in principle to a
Grothendieck site where the key idea is to replace sets with objects in a category.

Within the context of a category with pullbacks, the concepts of bundle and sheaf  are
equivalent. For example, one may consider the sheaf of sections of the bundle $\pi: E \rightarrow
G$ or those of the tangent bundle $TM$. Since we are working with noncommutative spaces, bundle
theory is no longer at our disposal and to build a geometry we will need to replace it with
something that will work without points and opens. It will also need to detect a non-local change
and allow us to replace tangent vectors with morphisms in a category.  Sheaf theory answers all of
these requirements. We build a sheaf out of $E$ using the modular structure of the fibres $E_g$,
and this will not be the same thing as the sheaf of sections of the Fell bundle because the base
space will be the space of objects `$B$' (see below) rather than $G$. Hence the new sheaf will not
be equivalent to $E$ as a bundle and its sectional algebra will not be isomorphic to $C^*(E)$.

First we clarify some notation and introduce the space of objects of $E$:

We denote `$B$' for base space the set of objects in a Fell bundle geometry and use the symbol $A_i$
for its smallest members with $i = 1,2,...,\vert G_0 \vert$. For a general member we write `$U$'.
When $G$ is discrete, we will treat $B$ as a space with the discrete topology. Let $A$ be
a semi-simple algebra: $A = A_1 \oplus A_2 \oplus A_3 \oplus A_4 \oplus...\oplus A_m$. Each
direct summand is an object in a Fell bundle geometry,  and we can take unions by direct summing
combinations of objects, so $A_1 \cup A_3 = A_1 \oplus A_3$ denoting this union by $A_{13}$,
$A_1 \cup A_{12} = A_{12}$,  $A_1 \subset A_{12}$ And $A_{123} \cap A_2 = A_2$. This is meant to be
a heuristic viewpoint, not a formal definition of a covering.

The process we call a `\textit{category theory switch-of-focus}' works in the following way. The
first step is deform the Riemannian manifold $M$ by replacing its tangent bundle with the
`tangential groupoid' $M \times M$ (\cite{dl}). To make the switch of focus, we replace each
groupoid object ($gg^* \in G_0$) with  the object $\pi^{-1}(gg^*)$ of $E$ and study the `space' of
objects $B$ instead of $M$. We then extend this to the replacement of each element $g$ of $G$ with
an element of the image of the *-functor $\pi^{-1}$ \cite{dfb}. Another way of putting this is that
we are replacing the sections of the local homeomorphism $d: G \rightarrow G_0$ with those of the
local homeomorphism $\rho: E \rightarrow E_0$ by taking objects to objects and morphisms to
morphisms. Note that this `category theory switch-of-focus' is not supposed to be any kind of
map at all because for each morphism in $G$ there are many choices of morphism in $E$. For example,
in the case of commutative $(E,D)$ we replace each point (or groupoid unit) with a whole copy of
$\mathbb{C}$. For more details see the explanatory remark after the following series of
definitions.

\begin{tangent field}[Tangent field and cotangent  field on $B$] Recalling that a tangent vector
field on a manifold $M$ is a smooth assignment of a tangent vector to each point in $M$, we define
a `\textit{tangent field}' on $B$ as a continuous field of morphisms in $E$, specifically, a
continuous assignment of a morphism of a Fell bundle geometry with domain $A_i$ of a Fell bundle
geometry $E$ to each object $A_i$ in $E$. Similarly, a \textit{cotangent field} is a continuous
assignment of a morphism of $E$ with range $A_i$ to each object $A_i$ in $E$. So each morphism in
$E$ defines a generalised `tangent vector'.

\begin{Comment:}
To have emphasized their algebraic nature we could have used the term `Fell bundle derivation' instead of
tangent field were it not for the fact that in noncommutative geometry the distinction between
$x$ and $a \rightarrow d_x$ (where $a$ is in the coordinate algebra, $x$ is a section and $d_x$ is the
derivation it defines) is important.
                \end{Comment:}

\begin{sections} All tangent fields are sections of $E$ but not all sections of $E$ are tangent fields.
\end{sections}

With these generalised (co)tangent vectors, we can formulate a `(co)tangent space'.

\end{tangent field}

\begin{tangent space}[Tangent space  and cotangent space at $A_i$.] The (co)tangent space at $A_i$
is the set of all (co)tangent fields on $B$ that have (range) domain $A_i$. \end{tangent space}

Below we use the term `sheaf' loosely as we are working outside
the field of point-set topology and $U$ is not an open set of points; we are exploiting
Grothendieck's principle that sheaves can be defined where points are replaced by objects in a
category.

\begin{tangent sheaf}[Tangent sheaf] A \textit{tangent sheaf} $sh_E$ is the following assignment of
tangent fields to each $U \in B$.


(To construct a sheaf we need two pieces of data and two axioms  (\cite{wiki}). Recall that $U$
denotes a member of $B$ and that any $U = A_{i,~j,..,~\vert G_0 \vert}$ is a union of smallest
members $A_i$.)

A tangent field on $A_{i,~j,..,~\vert G_0 \vert}$
is a choice of tangent field for each of those $A_i$s. (The set $sh_E(U)$ does not necessarily form
a group because in general morphisms are not added    together in the same way as vectors are
added.) The choices of tangent fields on the overlaps $U \cap V$ are to be equal. The second piece
of data is the restriction morphism. Let $V$ denote another member of $B$ that is contained in $U$.
The restriction morphism, $res_{V,U} : sh_E(U) \rightarrow sh_E(V)$ restricts the data of $sh_E(U)$
to that of $sh_E(V)$ such that the set $sh_E(U \vert_V)$ is precisely the set $sh_E(V)$.
\end{tangent sheaf}

To see that $sh_E$ is a sheaf observe that the restriction map satisfies the normalization axiom and
the gluing axiom. The normalization axiom states that $sh_E(U=0)$ is a one element set. Indeed,
there are no `tangent fields' on $U=0$, so the set $sh_E(U=0)$ is the one element set, the empty
set. The restriction axiom is satisfied by definition because it says that for $U$ a union of the
$A_i$, then an element of $sh_E(U)$ is the same as a choice of elements in $sh_E(A_i)$ for each
$i$, subject to the condition that those elements are equal on the overlaps $A_{ij..} \cap
A_{kl..}$ (\cite{wiki}).

\begin{cotangent sheaf}[Cotangent sheaf] We define  the cotangent sheaf by replacing each occurrence
of `tangent field' with `cotangent field' in the previous definition. Clearly, it is isomorphic to
$sh_E$.

\end{cotangent sheaf}

\begin{sheaf opp}[The sheaf $sh_{E^{opp}}$] This  sheaf is defined in the analogous way to $sh_E$.
In this case one simply replaces $A$ with $A^{opp}$ and constructs the corresponding sheaf in the
same way as $sh_E$ is constructed.

\end{sheaf opp}

 \begin{sheaf Cl}[Clifford sheaf $sh_{Cl}$] The sheaf $sh_{Cl}$ is defined in the analogous way to
$sh_E$. In this case one simply replaces $E^0$ with $Cl^0$ and  $A$ with the algebra of sections of
$Cl^0$ and constructs the corresponding sheaf in the same way as $sh_E$.
                                   \end{sheaf Cl}

\begin{comments on sheaf} As for any sheaf  constructed from sectional data, there is a local
homeomorphism $\pi : E \rightarrow B$ where $E$ is the \'etale space associated to the sheaf. $E$
consists of the disjoint unions of the stalks $sh_E(A_i)$. The stalk over $A_i$ is the tangent
space at $A_i$ and is identifiable with the set of all left $A_i$-modules (and the stalk over $A_i$
pertaining to the cotangent sheaf is identifiable with the set of all right $A_i$-modules) and
therefore, the union which is the whole \'etale space, with convenient notation, is the original
Fell bundle $E$. The categorical description usually given to \'etale spaces does not correspond to
that of the Fell bundle.

Since the \'etale  space is composed of modules, we see that the space $sh_E$ is a module over $B$,
and this is analogous to the fact that the set of sections of $TM$ form a module over
$C^{\infty}(M)$.

\end{comments on sheaf}

\begin{explanatory remark}[`category switch-of-focus']
                        In the preliminaries we recalled that the fluctuations of a Dirac operator
in a finite dimesional algebra $A_{\mathrm{finite}}$ involved unitary elements of
$A_{\mathrm{finite}}$ rather than having come from the process of lifting the algebra to
the spinors. For this reason a noncommutative Fell bundle geometry does not have a separate
Clifford algebra and it is also for this reason that the tangent fields involve the Clifford
algebra, and this is why the `category switch-of-focus' is not a well defined map, but a
qualitative process. See later that a Fell bundle Dirac operator will be a tangent field and in the
classical limit, this will turn out to be an element of $T^*M$ already contracted with an element
of the Clifford algebra.

 \end{explanatory remark}

\subsection{\textit{Observables}}

\subsubsection{\textit{Discussion on the noncommutative algebra of observables}}


Let $(E,D)$ be a Fell  bundle geometry. Note that the sections of $sh_E$ are closed
under multiplication but not under addition and therefore they do not generate a $C^*$-algebra. If
all elements of the sheaf are derivations then they form a Lie algebra with respect to the
commutator bracket. Under both multiplication and addition, the sections generate a representation
of $C^*(E)$ and we view this as the $C^*$-algebra of observables. We expect the set $sh_E$ is
isomorphic to the group of bisections of the general linear groupoid $GL(E)$.


Note also that the observables themselves are not bounded if $H$ is infinite dimensional: this is
the same as the usual scenario of noncommutating operators $ab-ba=1$; if $H$ is infinite
dimensional, this relation can only be satisfied for $a$, $b$ unbounded \cite{book} whereas the
algebra of observables $C^*(T^*M)$ is of course bounded.  To proceed, one may choose to restrict
the domain of $H$ to the intersection of the domains of definition of all the unbounded operators,
or exponentiate the unbounded operator to obtain a unitary. Obviously when $H$ is finite
dimensional, no such circumvention is necessary.

Observables should also be self-adjoint but the set of sections with these
properties do not form an algebraicly closed set. In contrast, the sections $x$, $y$ satisfying
$xJ=Jx$ do close:

\begin{equation}  \label{algebraicly closed set}
yJ = Jy,~~ xJ=Jx, ~~xJy=Jxy,~~(xy)J=J(xy)
\end{equation}


\subsubsection{\textit{Gauge invariance}}

We refer the reader to the short introduction to `fluctuations of the metric' given in the
preliminaries.

It is well known that  observables must be gauge invariant. The gauge invariant Dirac operator is
the covariant derivative formed by fluctuating it. We define a gauge invariant tangent field $x^f$
by:

\begin{equation} \label{x fluctuations}
   x^f = \sum_{\textrm{finite}} r_j  U_j x   U_j^{-1}
\end{equation}

which in the noncommutative $(E,D)$ case is for all $x \in sh_E$ and where $U_j$ denotes a unitary
element of the noncommutative algebra $A$, and in the commutative Fell bundle geometry case, $x \in
sh_{Cl}$ and $U_j$ provides the spinor lift $L$ of the automorphisms of $A$.

This question is especially relevant from the physics point of view when the Clifford bundle of a
Riemannian spin manifold is twisted with a bundle representing internal space, which in the
noncommutative standard model is a finite dimensional noncommutative algebra $A_{\mathrm{finite}}$
and can even be thought of as an extension of the Clifford algebra. The overall algebra of the
total space is a tensor product algebra and is sometimes called an `almost commutative' algebra
$C^{\infty}(M) \otimes A_{\mathrm{finite}}$ and we must fluctuate the Dirac operator on the first
factor space in the algebra of the second.

%



\subsubsection{\textit{Physical interpretations}}

Recall that the Dirac  operator occurs implicitly in the Hamiltonian as a square root of the
Laplacian. Above, the observables are built out of a set of sections from which one makes
the canonical choice for the Dirac operator. We give an explicit example of this below in 5.1 We
make the physical interpretation that the observables pertain to the force of Higgs gravity acting
on the fermions as represented by $H$. This comes from the noncommutative standard model where the
Higgs and the Dirac operator encode the connection and the metric respectively and the eigenvalues
of $D$ are directly related to the work done against the Higgs force in transporting a fermion
between fuzzy points in the noncommutative manifold. This is an open path in the space $E$, in
other words a morphism. The observables are not diffeomorphism invariant, as they shouldn't,
because the fermions live on the space, they are not the intrinisic geometric degrees of freedom
for gravity. A parallel transport along a closed path yields an element of the algebra $A$ and we
expect this to lead towards a diffeomorphism invariant algebra of observables for gravity on a Fell
bundle geometry.

Note that in the commutative case, our construction of a commutative Fell bundle geometry is in
harmony with the idea that categorification leads to quantization because it involves a
deformation of the cotangent bundle, whereas in the noncommutative case this deformation is
an intrinsic consequence of the noncommutativity and it is not obvious and probably not even true
that our categorification leads \emph{directly} to quantization of the particle mechanics on the
noncommutative manifold. We intend to continue this discussion on physical implications in a future
paper.

\subsection{\textit{Construction of $D$}}

In spectral triples, $D$ is defined  without any reference to the tangent bundle, so in a purely
algebraic sense without points. This makes them a sort of Gelfand Naimark counterpart for geometry.
In order to formulate an algebra of observables for a particle system on a noncommutative manifold,
we introduced an algebraic alternative to tangent bundle, namely tangent sheaf. The definition
below is inspired by the fact that given any Riemannian geometry, one can always make an explicit
construction of a Dirac operator by taking a section of the cotangent space and contracting with an
element of the Clifford algebra \cite{Berline}. This is the initial Dirac operator that one
generalises through the equivalence principle (see fluctuations section in the preliminaries).


Note that a consequence of this construction for physics is that we may
view the Dirac mass matrix, which in the noncommutative standard model encodes a parallel transport
between left and right, as a morphism in a $C^*$-category.

Dirac operators are first order differential operators and in noncommutative geometry, that is
usually taken to mean that $D$ defines an \emph{inner} derivation. An inner derivation on a Banach algebra $A$
is a derivation $\delta$ from $A$ into an $A$-bimodule $N$ for $x \in N$,
$a \mapsto [x,a] ~\forall a \in A$  (\cite{Harti}). The first order condition
is one of the noncommutative geometry axioms and it states that:

\begin{equation} \label{condition}
   [[D,a],b^{opp}] = 0, ~~\textrm{or}~~ [[D,b^{opp}],a] = 0
\end{equation}

Let $D$ define the inner derivation: $\delta : a \mapsto [D,a]$ and
$\delta : ab^{opp} \mapsto [D,ab^{opp}]~ \forall a \in A, b^{opp} \in A^{opp}$. Then the definition of derivation:
$\delta(ab^{opp}) = a \delta (b^{opp}) +  \delta( a)b^{opp}$ says that:

\begin{equation} \label{delta}
  D(a b^{opp}) - (a b^{opp})D = a(D b^{opp} - b^{opp} D)  +  (Da-aD)b^{opp}
\end{equation}

for all $a \in A$ and all $b^{opp} \in A^{opp}$. This provides the condition on $D$ and $\rho(A)$ stating
that $D$ defines a derivation that is inner. By substituting \ref{condition} into \ref{delta} and vice
versa we find that they are equivalent. This condition was originally included in the axioms as one on
the algebra for a given space of Dirac operators (referring to for example \cite{sap} or \cite{gravity}).
The opposite algebra is involved in order to carry the statement $d(fg)=d(gf)$ into
noncommutative geometry. That is, $\delta(b^{opp} a) = \delta (a b^{opp})$ because $a$ and $b^o$ commute.
Equivalently, $[[D,a],b^{opp}]=[[D,b^{opp}],a]$.

\begin{Fell bundle D}[Initial Fell bundle $D$] An initial
Dirac operator $D$ on a Fell bundle geometry is a section $x$ of the cotangent sheaf on $B$
tensored with a section of the cotangent sheaf dual to $sh_{E^{opp}}$ (take a morphism with range
at each object $A_i$ of $E$ and tensor it with a morphism with range at each $A_i^{opp} \in
E^{opp}$) such that $x=x^*=Jx^*J^*$. $D$ defines an inner derivation $a \mapsto [D,a] ~\forall a
\in A = C^*(E^0)$.

 
\end{Fell bundle D}

Below we give a separate definition of initial Dirac operator on a commutative Fell bundle geometry
for the sake of clarity but of course this is just a special case of the previous definition.

\begin{Dirac operator}[Commutative initial Fell bundle $D$] An initial Dirac operator $D$ on
a commutative Fell bundle geometry is a continuous field of morphisms of $Cl$ (or element of
$Sh_{Cl}$) that defines an inner derivation of $A$ such that $D=D^*=JDJ^*$.

 \end{Dirac operator}

These notions are analogous to an initial Dirac operator on a Riemannian manifold but we can
make the definition of Fell bundle Dirac operator more general by invoking the equivalence
principle, that is, by fluctuating it in the Clifford algebra.

\begin{General D}[Fell bundle Dirac operator] A \textit{general or `fluctuated' Fell bundle Dirac
operator} is given by the finite sum:

\begin{equation}
           D^f =   \sum_j r_j U_j D U_j^*; ~~~ U_j~ \mathrm{a~ unitary ~in}~ A,~ r_j \in \mathbb{R}
\end{equation}

where in the commutative case, $A$ stands for $Cl^0$ and in the noncommutative case, $A$ stands for
$C^*(E^0)$ and where $D$ is an initial Fell bundle Dirac operator.
        \end{General D}

The recipe where we `take  a morphism...' is demonstrated in example 5.1 below. We hope that the
reason why the opposite algebra and $E^{opp}$ is involved will become evident in the examples
section 5, where we explain a physical motivation for defining Fell bundle geometries.

\begin{not inner} Note that for noncommutative $(E,D)$, $D$ is an observable (see earlier section
on physical interpretations). Even after imposing the first order
condition on $\rho(A)$ so that it be satisfied for $D$, not all $x \in sh_E$ necessarily define
inner derivations such it be satisfied by $x$ for all $a \in A$. In that case those sections may
define other derivations that are not inner, for example, $\delta_x:A \rightarrow N,~~a \mapsto 0$.

\end{not inner}

\begin{junk} $D$ is defined in  terms of the Clifford algebra which is supposed to be isomorphic to
the differential algebra, but it is not and one has to divide out the `junk' forms. So far we have
left this out of this study but it is a point (raised by Bertozzini) to be addressed. Here we are
constructing $D$ more from the point of view of the differential algebra and perhaps derivations
that are not inner will be relevant to this discussion.

\end{junk}

\section{A categorification of real spectral triples}

According to Connes, a noncommutative manifold is a  spectral triple together with a set of axioms
\cite{gravity}. His reconstruction theorem confirms that the notion of real spectral triple is a
true generalisation of Riemannian (spin) manifolds to noncommutative geometry. Therefore, if a Fell
bundle geometry is a noncommutative manifold, it ought to be a real spectral triple. In this section
we demonstrate this to be the case and we determine what class of spectral triples can be repackaged
as Fell bundle geometries and can hence be viewed as $C^*$-categories, that is as categorified
not necessarily commutative geometries. This section can be summarised by statements (a) to (c):

\begin{main theorem} (a) All Fell bundle geometries (whether commutative or noncommutative) are real
spectral triples. (b) All commutative real spectral triples with unital, complex $C^*$-algebras are
the classical limit of some Fell bundle geometry.  (c) The class of noncommutative finite real
spectral triples with complex algebra are Fell bundle geometries are those that satisfy
additional constraints that arise as a corollary of (a). \end{main theorem}

\begin{Constraint corollary}[New constraints on $D$] The Dirac operator for a noncommutative
Fell bundle geometries satisfies all the conditions put upon it by the noncommutative geometry
axioms but the definition of Fell bundle Dirac operator constrains it even further. The set of
spectral triples that are Fell bundle geometries will not include those that do not satisfy these
new constraints. In a further paper emphasising physical implications we aim to investigate whether
this implies eliminating only unphysical spectral triples.

A consequence of the
interpretation of the Dirac operator as a field of morphisms in a category and from a section in
the cotangent sheaf is that it be constrained in the following way. \textit{This is that $D$ is to
be a tensor product of two matrices} and these two matrices are in the hom-sets, the imprimitivity
bimodules over the respective object algebras. A further constraint is afforded to $D$ also because
it is constructed from the category theory concept of tangent field. This is shown explicitly by
example in section 5, \textit{removing `leptoquarks', which are extra unphysical degrees of freedom}
that can over-constrain the equations of motion (\cite{smv}).          \end{Constraint corollary}

        \section{Example}

\begin{two point space}[The two point space] Consider a universe consisting of two identical 4
dimensional manifolds or ``sheets'', one labelled ``L'' and the other ``R''. The Higgs acts as a
connection defining parallel transports between the two sheets. The algebra describing the discrete
space of two points is $\mathbb{C} \oplus \mathbb{C}$. Due to fermion doubling (or quadrupling), we
will actually work with four points to write down the spectral triple: $(A, H=\mathbb{C}^4, D,
\chi, J)$ where $A=\mathbb{C} \oplus \mathbb{C} \oplus \mathbb{C} \oplus \mathbb{C}$, $\chi=
\mathrm{diag}(1, -1, 1, -1)$ and $D$ is the 4 by 4 matrix:

\begin{equation}     \label{D two point space}
D=\left(   \begin{array}{cccc}
0         &         \bar{m}    &   0        &   0\\
m         &          0          &   0        &   0\\
0         &          0          &   0        &  m\\
0         &          0          &   \bar{m}  &  0
\end{array}  \right)
\end{equation}

$m \in \mathbb{C}$, with $\bar{}$ denoting complex conjugation. $D$ acts on $H$ of which a typical
element we denote $\Psi=(\psi_L, \psi_R, \psi_{\bar{L}}, \psi_{\bar{R}})^T$.  The algebra $A
\otimes A^{opp}= \mathbb{C} \otimes \mathbb{C} ~~\oplus ~~\mathbb{C} \otimes \mathbb{C} ~~\oplus~~
\mathbb{C} \otimes \mathbb{C} ~~\oplus ~~\mathbb{C} \otimes \mathbb{C}$ has a faithful action
$\rho$ on $H$ with $A$ acting on the left and $A^{opp}$ on the right. $J$ acts on $H$ in the
following way: $J(\psi_1 ~~ \bar{\psi_2})^T = (\psi_2 ~~ \bar{\psi_1})^T$. This spectral triple is
even since $[a, \chi]=0$, $\chi^2=1$ $D \chi = -\chi D$.

To proceed with the Fell bundle categorification, we consider the product bundle $(E \otimes E^{opp}, \pi, G)$ where $G$ is the pair groupoid on the discrete space consisting of the 4 points $G$ = Pair(4). The groupoid algebra is identified with the algebra of sections of $E$ and that is the set of 4 by 4 matrices. The fibres (all copies of $\mathbb{C}$) over $G$ are Morita equivalence bimodules over the objects and their elements are the morphisms of $E$.  The objects of the categorified real spectral
triple are  the direct summands of the algebra $A \otimes A^{opp}$ and these define the objects of
$E \otimes E^{opp}$. The morphisms in the categorification are the elements in the fibres over $G$.
This is a $C^*$-category and can be represented on Hilb. The base space in the generalised tangent
bundle or tangent sheaf is $B=\pi^{-1}(G_0)$, which is the set of 4 objects in $E \otimes E^{opp}$
and we treat it as a discrete topological space.

Figure 1 presents the 4 objects of $E$:-

\vspace{0.5cm}
\begin{figure}[ht]   \label{d}
\qquad \qquad \qquad \qquad
\xymatrix@R=4pc@C=4pc{
 \mathbb{C} \ar@{-}[d] \ar@{}[r]  &    \mathbb{C} \ar@{-}[d] \ar@{}[r] &   \mathbb{C}\ar@{}[r] \ar@{-}[d]  &    \mathbb{C} \ar@{-}[d] \\
L  \ar@{}[r] &  R \ar@{}[r] &  \bar{L} \ar@{}[r]   &    \bar{R}    }

\caption{The space of objects $B$}
\end{figure}
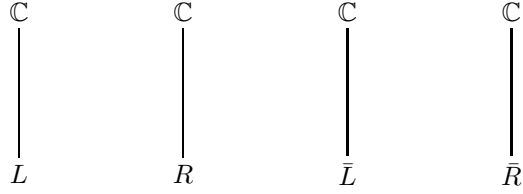

\vspace{0.5cm}

where we have labelled the objects of $G$ as $G_0 = \{L, R, \bar{L}, \bar{R} \}$.

With figure 1 in mind, we express $\rho(A)$ as matrices of the form:

\begin{equation}    \label{rho}
\rho(a) =
\left( \begin{array}{cccc}
\rho_L   &       0     &       0       &     0  \\
0        &    \rho_R   &       0       &     0  \\
0        &      0      &    \rho_{\bar{L}}   &     0  \\
0        &      0      &       0       &  \rho_{\bar{R}}
\end{array}   \right)
\end{equation}

for all $a \in A$, with basis indexed by $G_0$.

And $\rho(A^{opp})$ is given by the matrices $\rho(b^{opp}) := J \rho(b)^* J^* \in \mathbb{C} \oplus \mathbb{C} \oplus \mathbb{C} \oplus \mathbb{C}$ for all $b \in A$ which take the form:

\begin{equation}    \label{rho opp}
\rho(b^{opp}):=
\left( \begin{array}{cccc}
\rho_{\bar{L}}   &       0     &       0       &     0  \\
0        &    \rho_{\bar{R}}   &       0       &     0  \\
0        &      0      &    \rho_L   &     0  \\
0        &      0      &       0       &  \rho_R
\end{array}   \right)
\end{equation}

To finish  building the Fell bundle geometry and check that $D$ can be inferred from it, we
construct the cotangent sheaf. Seeing as $\mathbb{C} \otimes \mathbb{C} \cong \mathbb{C}$, the
product bundle is isomorphic to $E$, so we only need to work with $E$. Using the definition we gave
for cotangent field together with the condition that they commute with $J$, the cotangent sheaf
comprises matrices of the 4 following forms.

The two  diagrams represent the choice of a morphism at each object and are positioned underneath
the matrix representation of the same tangent field. Arrows are morphisms of $E$ (not $G$) and
points in the diagram represent objects of $E$.

\begin{equation}     \label{canonical choice}
\left(   \begin{array}{cccc}
0         &         \bar{m}    &   0        &   0\\
m         &          0          &   0        &   0\\
0         &          0          &   0        &  m\\
0         &          0          &   \bar{m}  &  0
\end{array}  \right)
\end{equation}

\vspace{0.5cm}
\begin{figure}[ht]
\qquad \qquad \qquad \qquad \quad
\xymatrix@R=4pc@C=4pc{
  {\bullet}  \ar @ /^1pc/ [r]    &   {\bullet}  \ar@ /^1pc/  [l]     &   {\bullet}  \ar @ /^1pc/ [r]   &   {\bullet} \ar@ /^1pc/  [l] }

 \end{figure}
\vspace{0.5cm}

\begin{equation}
\left(   \begin{array}{cccc}
0         &          0          &    g        &   0\\
0         &          0          &   0        &   h\\
\bar{g}   &          0          &   0        &  0\\
0         &          \bar{h}    &   0        &  0
\end{array}  \right)
\end{equation}

\begin{equation}  \label{w}
\left(   \begin{array}{cccc}
w         &          0          &   0        &   0\\
0         &          z          &   0        &   0\\
0         &          0          &   \bar{w}  &  0\\
0         &          0          &   0        &  \bar{z}
\end{array}  \right)
\end{equation}

\vspace{0.5cm}
\begin{figure}[ht]
\qquad \qquad \qquad \qquad \quad
\xymatrix@R=4pc@C=4pc{
  {\bullet}    \ar @(ul,dl)   &   {\bullet}  \ar @(ul,dl)     &   {\bullet}  \ar @(ul,dl)   &   {\bullet} \ar @(ul,dl) }

\end{figure}

\vspace{0.5cm}

\begin{equation}  \label{y}
\left(   \begin{array}{cccc}
0         &          0          &   0        &     y \\
0         &          0          &   y        &   0\\
0         &          \bar{y}    &   0        &  0\\
\bar{y}   &          0          &   0        &  0
\end{array}  \right)
\end{equation}

$x,y,g,h,w,z \in \mathbb{C}$. Note  that the tangent fields are obtained by reversing the direction
of each arrow, and that it is therefore obvious that the tangent sheaf and the cotangent sheaf are
isomorphic. Also note that if the orientation of the manifold is switched then the cotangent and
tangent sheaf are exchanged.

The self-adjoint cotangent  field \ref{canonical choice} is the Dirac operator $D$. It is the only
choice from \ref{canonical choice} to \ref{y} that both defines an inner derivation and
anticommutes with $\chi$ so we do not actually need to make a canonical choice for $D$, because
there is only one answer.  The first of the two diagrams demonstrates that except for in the case
of \ref{w}, \textit{ tangent field morphisms ``know'' whether they are ranged or sourced at a given
object; that means that the orientation of the morphisms is unambiguous and the manifold is
orientable}. This illustrates the fact that the matrix \ref{canonical choice}
satisfies $D \chi = -\chi D$. \textit{Note that there are no additional degrees of freedom appearing
in the Dirac operator (known as ``leptoquarks'')}: this is the first constraint on $D$ that appears
in the current formulation but that does not come from the real spectral triple axioms; what we
have is that the \textit{$S^o$-reality condition is automatically satisfied}. Since $\mathbb{C}
\otimes \mathbb{C} \cong \mathbb{C}$ there is no further constraint on $D$ for this example.

\vspace{0.5cm}

\begin{figure}[ht]  \label{d local}
\qquad \qquad \qquad \qquad \qquad \qquad
\xymatrix@R=6pc@C=1pc{
 \rho_L   \ar[d]^{~\mathbb{C} ~~~~\otimes ~~~~\mathbb{C}}            \ar@{}[r]  & \otimes &  \rho_L^{opp}   \ar[d]^{~~~~~~\therefore~ D|_{\psi_L)} \in \mathbb{C} \otimes \mathbb{C}}  \\
\rho_R   & \otimes  &  \rho_R^{opp}   }

\caption{Local $D$ as a map from $\psi_L$ to $\psi_R$} \end{figure}

\vspace{0.5cm}

Figure 2 represents a morphism in a cotangent field together with its source and range objects.

We showed  earlier that the set of cotangent fields $x$ satisfying $xJ=Jx$ is algebraicly closed
(\ref{algebraicly closed set}), but one may also demonstrate this explicitly for this example by
(i) multiplying the matrices  together (\ref{canonical choice} to \ref{y}) or (ii) drawing diagrams
corresponding to all 4 types of cotangent field and then composing their arrows. This set of tangent
fields comprises a gauge invariant $C^*$-algebra.

\end{two point space}

\section*{Other applications}

Having a noncommutative  manifold with an explicit notion of generalised tangent bundle might
provide an alternative way to study the geometrical properties of these spaces, for example,
perhaps a connection could be defined as a horizontal distribution and alternative measures of
curvature could be studied for noncommutative spaces in analogy with those for commutative
manifolds that rely on the tangent bundle, for example departure from integrability and
infinitesimal holonomy loops. These constructions might lead to a new way of studying
non-local noncommutative geometry and also $C^*$-dynamical systems.

Since the space of charges is what internal space
is supposed to capture and not only chirality, a construction of double Fell bundle geometries
ought to be a more correct approach to modelling noncommutative spaces (\cite{dfb}). In that case,
the Dirac operator will be a field of 2-morphisms. This might also lead to applications in
noncommutative Yang-Mills theories.

Since the arrows in Feynman diagrams can be thought of as morphisms in a
category and spin foams are built using this idea, there might be applications to noncommutative
quantum gravity where the 4-manifolds are replaced by real spectral triples as double Fell bundles
with the boundary manifolds as the vertical category. Since $(E,D)$ is a $C^*$-category it has a
representation on Hilb \cite{LRG} and this fact could lead to a noncommutative topological
quantum field theory.

\section*{Acknowledgements}

Advice was gratefully received from Pedro Resende, Ryszard Nest, Paolo Bertozzini
and Ali Chamseddine. The affiliation is given on the title page.

\end{document}